\documentclass[preprint,1p,times,authoryear]{elsarticle}
\usepackage[pdftex]{color}
\usepackage[font=footnotesize,labelfont=bf]{caption}
\usepackage[font=footnotesize,labelfont=bf]{subcaption}
\usepackage{amsmath,amssymb,amsfonts}
\usepackage{algorithmic}
\usepackage{graphicx}
\usepackage{booktabs}
\usepackage{textcomp}
\usepackage{xcolor}
\usepackage{comment}
\def\BibTeX{{\rm B\kern-.05em{\sc i\kern-.025em b}\kern-.08em
    T\kern-.1667em\lower.7ex\hbox{E}\kern-.125emX}}
    
\PassOptionsToPackage{hyphens}{url}\usepackage{hyperref}    
\usepackage{color}
\newcommand{\todo}[1]{}
\renewcommand{\todo}[1]{{\color{red} TODO: {#1}}}
\usepackage{lineno}

\AtBeginDocument{%
  \providecommand\BibTeX{{%
    \normalfont B\kern-0.5em{\scshape i\kern-0.25em b}\kern-0.8em\TeX}}}

\journal{Journal of Systems and Software}

\begin{document}

\begin{frontmatter}


\title{Emotimonitor: A Trello Power-Up to Capture Emotions of Agile Teams}


\author{Mohammed-Amr Abd El-Migid\corref{label1}}

\author{Damon Cai\corref{label1}}

\author{Thomas Niven\corref{label1}}

\author{Jeffrey Vo\corref{label1}}

\author{Kashumi Madampe}

\author{John Grundy}

\author{Rashina Hoda}

\address{Department of Software Systems and Cybersecurity, Monash University, Melbourne, Australia.}

\cortext[label1]{Equally contributed as first authors}

\begin{abstract}
  In recent years, Agile methods have continued to grow into a popular means of modulating team productivity, even garnering a presence in non-software development related industries. The uptake of Agile methods has been driven by their flexibility, making them more suitable for many teams when compared to traditional approaches. However, an inevitable expectation for an Agile workflow is a higher level of change and uncertainty regarding requirements and tasks, which can ultimately have impacts on team member emotional states. The extent of such emotion impacts has motivated our research into the manner in which emotional states evolve in an Agile setting, along with whether such emotions can be accurately measured. To this end, we have developed \textit{\textit{Emotimonitor}}, a Trello power-up designed to capture information on emotions of team members as they relate to their technical tasks through a user-friendly interface. \textit{\textit{Emotimonitor}} will better enable team members to express their emotional states through emoji reactions on Trello cards, while also providing team leaders with a dashboard summarising these reactions as visualisations and statistical data. It is extensible and potentially provides an outlet for team members operating in Agile environments to better express their emotional states. 

\end{abstract}

\begin{highlights}
\item Real-time capturing of emotional responses shown to tasks in Trello cards is possible
\item We developed a prototype of a Trello power-up which captures, and evaluates the emotional responses shown to tasks in cards of Trello boards
\end{highlights}


\end{frontmatter}

\section{Introduction}
\label{sec: Intro}
With the growing popularity of agile methods, teams have adopted innovative approaches to ensure productivity and facilitate a development cycle that is flexible and constantly evolving. With such an environment in which team members are required to welcome requirement changes late in development\citep{beck_manifesto_2001} among other things, examining the subsequent impact on emotions provides an opportunity to explore whether such emotions can be accurately captured and acknowledged to boost productivity. This is fundamental, as research indicates that emotions have direct linkages to productivity \citep{graziotin_happy_2014, graziotin_feelings_2015}, through stimulating cognitive processing abilities and creativity. Research has found the influence of emotions on the effectiveness of software engineers as “unquestionable” \citep{kolakowska_emotion_2013}. This, along with the current lack of support available for capturing developer emotions during development processes, necessitates an efficient emotion capturing tool. 

Trello is a widely used tool by Agile teams in the industry due to its Scrum and Kanban like conducive interface, lending itself to project management and task organization \citep{johnson_trello_2017}. The simple ability to create cards to represent tasks as well as to update the state of these cards by moving them between lists mirrors the aim of Agile methods such as Scrum in quickly developing, testing and releasing features which can be represented as moving cards between lists to update their state.

After considering existing alternatives, we have designed and developed \textit{Emotimonitor}, a Trello power-up (i.e. extension or plugin) that enables Agile team members to self-report their emotional reactions with respect to tasks. Such a tool relies on emojis to reflect emotional states and provides a platform for team leaders to view the responses of their team in order to justify taking actions such as reevaluating the requirements of tasks should there be a negative response to them. \textit{Emotimonitor} is extensible and has potential to improve team dynamics across agile teams, through enabling emotion identification to be a central element of retrospective processes.

The paper is structured as follows. Section \ref{sec:motivation} outlines the motivation for our work. This will be followed by Section \ref{sec:mt} which describes the research approach and development methodology used. Section \ref{sec:design} then discusses \textit{Emotimonitor}'s architecture followed by a discussion of its intended usage in Section \ref{sec:usage}. Section \ref{sec:evaluation} will describe our evaluation process, followed by Section \ref{sec:tv} discussing threats to the validity of our research along with Section \ref{sec:fw} which will outline planned future work for \textit{Emotimonitor}. This will lastly be followed with an analysis of related tools in Section \ref{sec:rw} and a conclusion of our research findings. 

\section{Motivation}
\label{sec:motivation}
Agile practices in project management are becoming increasingly prominent, not only within the software development industry in which they arose, but in other disciplines as well. They are distinguished from older “heavy-weight” methods primarily by the degree to which they encourage flexibility and evolution, both in responding to change and in driving change \citep{hoda_rise_2018} –  the Manifesto for Agile Software Development \citep{beck_manifesto_2001} fundamental to the Agile movement, calls for practitioners to ``welcome changing requirements, even late in development'', and describes an ideal Agile team as “self-organising”, one that "reflects on how to become more effective, then tunes and adjusts its behavior accordingly” \citep{beck_manifesto_2001}. During Agile projects, the expectations and emotional states of team members may be impacted in meaningful ways by the higher level of change and uncertainty – in tasks, expectations and environment \citep{hoda_rise_2018} – engendered by Agile practices, particularly in contrast to traditional practices such as the waterfall model \citep{madampe_towards_2020}. The extent of this impact on team members became our primary justification when examining how and to what extent these emotional states are likely to evolve in an Agile setting and how using this information may enable teams to better self-organise.

Emotions may have a fundamental impact on an Agile team’s ability to maintain productivity and sustain a healthy working environment \citep{graziotin_software_2014, graziotin_feelings_2015}. As such, the implications of not being able to acknowledge emotional states in order to actualise improvements in team processes is the main basis for our development of \textit{Emotimonitor}. We wanted to answer the following research question regarding the efficiency of \textit{Emotimonitor}:

\textbf{Can developers' emotional responses to changes in project requirements be captured effectively by extending project management tools, such as Trello?}

This research question motivated our development of a power-up for the project management application Trello, widely used across different Agile teams to organise tasks for team members. This power-up, \textit{Emotimonitor}, aims to allow team members to quickly and easily self-report their emotional reactions to an item of work (represented as a card) using a predefined set of discrete emotion states (represented as emoji). Team leaders are able to review these reactions both for a single item and across a project. The project-wide view can highlight correlations between particular factors and emotion states. Individual team members are also able to view their emotional reactions across different cards. We anticipate that \textit{Emotimonitor} will shed great insight into the relationship between Agile processes and the evolution of emotional responses in a team environment. Such insight will provide a future avenue to explore more engaging ways in which emotions can be directly elicited from team members. This may lead to more refined tools and power-ups and prompt further research into how teams can become better equipped to respond to evolving emotional states.Consider a situation where a group of developers are collaborating on a new project; a project manager provides a developer with new change requirements due the following day. With \textit{Emotimonitor}, the developer will be able to express their frustrations on the corresponding Trello card. Without it, the developer would lack a means of expressing their emotional state, preventing meaningful improvements in team processes from occurring. 

\section{Our Approach}
\label{sec:mtd}
Our aim was to develop a Trello power-up aimed at effectively capturing emotions of a team member at various stages of an agile project. Such a power-up would necessitate an intuitive, simple and fast user interface as well as the ability to capture the necessary data required. This information would include: The emotions of a team member with respect to a given card, The time of the emotional response, and the stage of the card when the emotional response was captured. The envisioned tool included a dashboard feature that would enable team managers to view the collective emotional responses of their team in a summarised manner. The intention was that managers would be able to leverage such summary statistics in order to actualise improvements on how a team operates along with addressing any visible issues.

\subsection{Choice of Emotion Schema}
We had to determine the most effective manner of eliciting emotional responses from users. A crtitical aspect of this was deciding on the emotion schema to be used in order to record team member responses. This schema would govern which emotions would be available for users to select with respect to Trello cards. Many such emotion schemas have been developed over many decades in psychology \citep{curumsing2019emotion}. We ultimately selected the discrete emotion schema defined by Harmon-Jones et al. \citep{harmon-jones_discrete_2016}. This consists of “eight distinct state emotions: anger, disgust, fear, anxiety, sadness, happiness, relaxation, and desire”  \citep{harmon-jones_discrete_2016}. This schema was chosen as the eight emotions were selected to capture distinctions along the dimensions of valence (positive/negative), arousal (high/low) and motivational tendency (approach/withdrawal), as well as other characteristics such as that do not lend themselves easily to a dimensional approach. 

\subsection{Prototyping Methodology}
Incremental development of a proof-of-concept for \textit{Emotimonitor} itself using an agile approach was fundamental to our project. Its development was iterative and involved frequent demonstrations to the 5th and 6th authors, who acted as product owners in the development process. The implementation of a basic user interface was the target for the first half of the project, the goal being to develop a primitive means of adding emotional responses to Trello cards. This cycle of development had an emphasis on user-friendliness, aesthetics, and ensuring emotions could be added to any card. The second half of the project was focused on implementing a data-storage solution and finalising the management dashboard/summary view containing statistics for managers to leverage. 

\subsection{Prototype Evaluation}
We wanted to evaluate our prototype with "real" users. After approval of our Human Ethics Committee, we recruited an agile development team to use \textit{Emotimonitor} and comment on its support for individual and management emotion capture and analysis.

\section{Design and Prototype}
\label{sec:design}

\subsection{Interface Design}

\begin{figure}[]
    \centering
    \begin{minipage}{0.3\textwidth}
        \centering
        \includegraphics[width=0.9\textwidth]{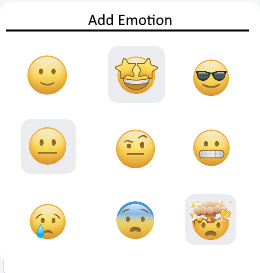} 
        \caption{First Mockup}
        \label{fig:first mockup}
    \end{minipage}\hfill
    \begin{minipage}{0.3\textwidth}
        \centering
        \includegraphics[width=0.9\textwidth]{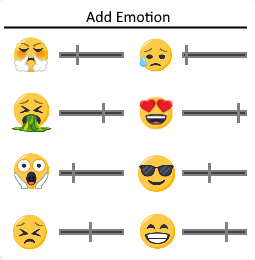} 
        \caption{Second Mockup}
        \label{fig:second mockup}
    \end{minipage}\hfill
    \begin{minipage}{0.3\textwidth}
        \centering
        \includegraphics[width=0.9\textwidth]{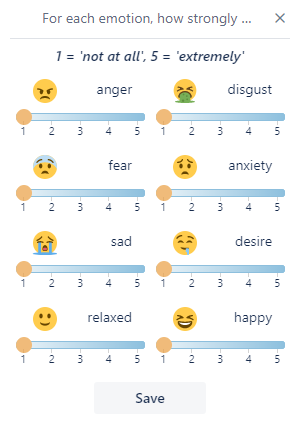} 
        \caption{Final Design}
        \label{fig:final design}
    \end{minipage}
\end{figure}

The development process began with designing an interface for \textit{Emotimonitor}; to go about this, a series of mock-up diagrams were fabricated by the 1st, 2nd, 3rd and 4th authors and reviewed by the 5th and 6th authors. Feedback from the latter culminated in an iterative improvement process where intended UI options and design were continuously refined and reviewed, before a final design was realised. Design considerations discussed as part of these meetings included the location of emojis, along with how scales could be incorporated in a manner that was both aesthetically pleasing and facilitated effective capture of reactions. Research conducted involved examining the most effective strategies in which emotions could be captured and contextualised, such as evaluating the nature of the emotional scale to be implemented and the necessity of subscales within each emotion. As such, a significant portion of the development process was dedicated to experimenting and critically evaluating different UI options that can foster easy emotion capturing (Figures \ref{fig:first mockup}, \ref{fig:second mockup}, \ref{fig:final design}). This process continued throughout the development of \textit{Emotimonitor}; one change that occurred relatively late in development was the introduction of the text above the sliders in Figure \ref{fig:final design} explaining the nature of the rating scale used.

\subsection{Prototype Implementation}
There are two major components that comprise \textit{Emotimonitor}. The first is a web app hosted by our team  that serves a Trello power-up embedded within the Trello web app as IFrames; the second is a backend consisting of a PostgreSQL database and accompanying REST API used to perform operations on the captured emotion data.

\textbf{Web App}
The web app is built using Preact\footnote{https://preactjs.com/}, a JavaScript library designed for small size and speed that is compatible with the widely used React\footnote{https://reactjs.org/} library. Integration with the Trello web app is accomplished via the Trello power-up client library. The app also makes calls to our accompanying API backend in order to store and retrieve the necessary data for capturing and visualising emotions. The app was prototyped as a server-side Express\footnote{https://expressjs.com/} app, but Preact was deemed to be more suitable due to its React compatibility and access to the vast React ecosystem. 

During the prototype phase the app was hosted using Glitch\footnote{https://glitch.com/}, which provides free dynamic hosting for Node.js web apps alongside a live code editor; this allowed us to iterate quickly in the early phases of development. As \textit{Emotimonitor}'s scope grew we eventually found Glitch to be too limiting due to the lack of support for build scripting, which precluded us from leveraging most of the React/Preact ecosystem. To rectify this, we shifted to a static web app architecture where our web app is compiled and served from a static server; we also moved our hosting onto Netlify \footnote{https://www.netlify.com/}, a static hosting service that provides automatic deployments and built-in HTTPS (necessary for our web app to be embedded in Trello) through a free plan.

\textbf{API and Database Backend}
Although Trello provides built-in data storage that would minimise required infrastructure and security concerns, we deemed this to be insufficient for our app’s requirements. As per Trello’s Power-Up Client Library documentation\footnote{https://developer.atlassian.com/cloud/trello/power-ups/client-library/getting-and-setting-data/}, plugin data stored on cards must be stringified and limited to a size of 4096 characters per ‘visibility scope’ (visibility scope being either ‘private’, i.e. the data is available solely to the member who set the data, or ‘shared’ i.e. visible to all who can view the card/board to which the data is attached). As neither scope satisfied the requirements for the data to be available to both the person setting the data as well as a chosen ‘manager’, as well as the data size limitation, storing data within Trello was deemed unviable.

PostgreSQL \citep{group_postgresql_2020} was chosen as the database engine as members of the team were already experienced with it, alongside its open-source nature being favourable for development. However, considering the simple Database schema utilized, a SQL database is not a necessity and future work could consider more efficient and/or simplified options.

The Amazon Web Services (AWS) Relational Database Service (RDS) was chosen to host the database, due to the available free tier allowing free development for this proof-of-concept as well as our team having greater familiarity with AWS compared to other cloud platforms. To interact with the database, AWS Lambda functions accessible via an API Gateway are utilized as HTTP endpoints from which to GET or POST data from/to the database. These endpoints require JWT (JSON Web Token) authentication in order to verify that requests originate from our Power-Up on Trello and are made by actual Trello users. JWTs are generated on the client by the Trello Client Library\footnote{https://developer.atlassian.com/cloud/trello/power-ups/client-library/t-jwt/}. There is further authentication for the chosen ‘managers’ who will have access to the data of the other team members, achieved via querying the Trello REST API on the server side to verify their status as an admin of the Trello board (this being the criterion for manager status).

\begin{figure}[]
    \centering
    \caption{Example card (A)}
    \label{fig:example card}
    \includegraphics[width=8cm]{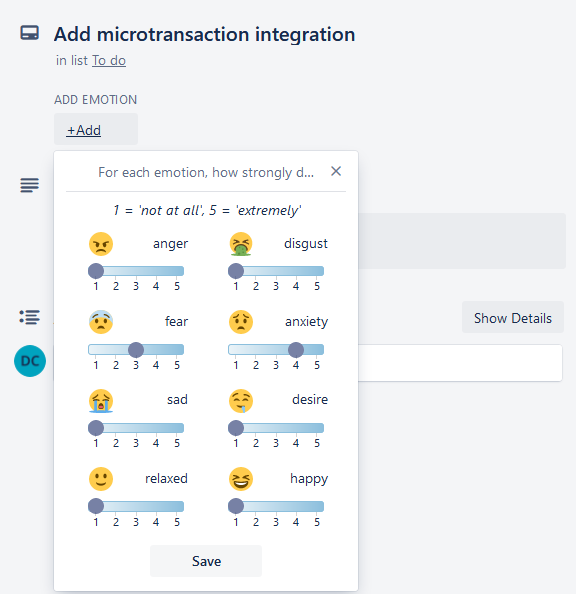}
\end{figure}

\begin{figure*}[]
    \centering
    \caption{Example Dashboard (B)}
    \label{fig:example dashboard}
    \includegraphics[width=16cm]{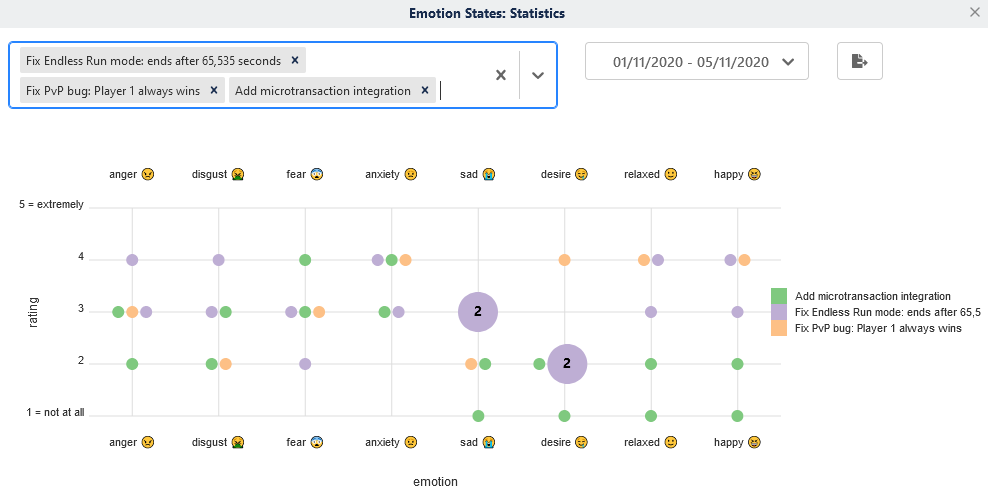}
\end{figure*}

\section{Usage Example}
\label{sec:usage}

The primary intention of \textit{Emotimonitor} is to assist managers in evaluating the emotional health of their team members, and for team members to self-reflect on their emotions when handling requirements changes. As outlined previously, the constantly changing Agile landscape can result in new requirement changes which can have an impact on team morale and individual feelings \citep{madampe_towards_2020}. By allowing managers to leverage key information about their team's emotional state via the manager dashboard, we intend for managers to be better equipped to recognise and resolve issues in the team. A notable testament of this would be the identification of a team member continuously displaying a notably strong ‘sad’ reaction to various cards on the board; a manager would be able to use this to justify conversations with the team member concerning their mental health and productivity. 

Beyond the identification of silo-ed team member sentiments impacting on productivity, \textit{Emotimonitor} aims to enable the tracking of team performance over the course of various sprints. With respect to this, preliminary research team discussions revolved around the topic of privacy; that is, should team members be capable of viewing each other’s reactions on tickets. A potential disadvantage of complete transparency here would be team members' potential apprehension about being truthful in their responses; that is, team members may be less willing to provide an honest depiction of their emotional state lest they be judged by their teammates. Given that \textit{Emotimonitor} is intended to be utilised further by teams, an anonymous approach was adopted. 

When considering Agile development practices, \textit{Emotimonitor} facilitates retrospective sessions by better enabling managers to voice potential issues to their team, thereby increasing productivity. While adhering to privacy constraints, managers can provide meaningful information to their team during these sessions and foster further conversations regarding why certain emotions are being provoked for certain tasks. For instance, a task receiving an abundance of negative emotions would prompt a manager to engage in discourse regarding how a team feels about a task and whether these feelings can be remedied through future actions e.g. training, increased communication, process changes. 

Recall the dilemma outlined in Section 2. Consider a developer, Kashumi, who is working on a free-to-play mobile game and feeling somewhat anxious about integrating microtransactions into the game: \textit{Emotimonitor} provides a means of capturing her emotional state not currently possible in Trello. By navigating to the card in question and selecting the button under 'Add Emotion', Kashumi will be provided with the following set of emotional scales, where she can select relevant emotions that apply to her such as 'anxiety' (4) and 'fear' (3). (See A: Figure \ref{fig:example card}).

After Kashumi and the other developers enter their emotional responses, her project manager Rashina can accurately gauge the emotional health of their team through a management dashboard (See B: Figure \ref{fig:example dashboard}). The dashboard  contains statistics regarding how the emotions of certain cards have changed over time, and would enable Rashina to discern whether team members have emotional states that ought to be addressed. In the above example, Kashumi's anxiety and fear about the microtransaction integration card are visible as a peak on the dashboard chart, which allows Rashina to identify and acknowledge the card as a potential issue, fostering greater productivity and an improved team dynamic.

\section{Evaluation}
\label{sec:evaluation}

\subsection{Evaluation Method}

In preparation for user testing, ethics approval was taken from the Monash University Human Research Ethics Committee \footnote{Approved by Monash Human Research Ethics Committee. Approval Number: 26387}.  When considering the evaluation plan, it was planned for volunteer Agile teams currently enrolled in a Bachelor of Software Engineering (Honours) at Monash University undertaking Software Engineering Research Project (FIT4003 course) to be part of the trial after request. A single trial of \textit{Emotimonitor} using team member satisfaction as a proxy for its effectiveness was undertaken. 

An Agile team consisting of 4 individuals  was provided with a dummy Trello board with the \textit{Emotimonitor} power-up added. All members of the team but one would be allocated ‘Team Members’ while the remaining member would be a ‘Manager’. ‘Team Members’ were allocated and assigned dummy tickets on the aforementioned board in advance. They would then be invited to view the board and assume the persona of an individual operating in an Agile team. ‘Team Members would be prompted to enter emotional responses on both assigned and unassigned tickets prior to a predefined deadline, before then being prompted to complete an evaluation form describing their experience with \textit{Emotimonitor}. This form enabled us to obtain insight into the usability of the application, a rating of its aesthetics and recommendations on potential improvements and refinements. 

Following the completion of ‘Team Member’ evaluations, the ‘Manager’ would be prompted to visit the dashboard and examine the management dashboard / summary view. Managers would be encouraged to examine the information present before completing a separate evaluation form outlining their experience. The goal would be for ‘Managers’ to share whether they found the information pertaining to the emotional responses useful, insightful and indicative of ‘supposed’ underlying issues; if this is the case, we could consider the evaluation successful.

\subsection{Evaluation Results}
Four participants were involved in our formal evaluations. When considering a ‘Team Member’s’ perspective, although overall experience with \textit{Emotimonitor} was found to be somewhat positive, the overall user-friendliness of the user interface was not as well-received. Despite this, users felt that it was capable of properly reflecting their emotions to some degree. However, several users expressed concerns that certain emotions pertaining to the chosen emotion schema lacked relevance for an Agile software development team, the prime example being the emotion ‘disgust’. 

Recorded positive aspects included \textit{Emotimonitor}: Being innovative and new, Being simple to access, Being intuitive to use, and Having scaled options of emotions enabled a good variety of potential responses.

However, common weaknesses identified included: There being too many emotions available which created confusion, The fact that some cards already reflect specific changes in requirements; this would thereby result in attached emotions not actually capturing a proper change in emotions, The lack of a feedback prompt upon saving responses, and Possible redundancies if opposing emotions are selected, such as ‘happy’ and ‘sad’.  When considering the ‘Manager’s’ dashboard, some criticism was directed at the clarity of the statistics presented on the management dashboard, along with how insightful such data might actually be in an Agile team context. 

Comments included that the manner in which statistics are presented made it somewhat difficult to gauge overall emotion states, most notably due to the usage of a grid view. A recommendation from a test user was that data be presented in the form of averages whilst utilising the same interface, as this would offer more understandable data at a glance. Another recommendation was also being able to view key ratings on cards in contrast to having managers rely purely on the management dashboard. This would facilitate an analysis process and save managers the time-consuming and tedious process of going back and forth between cards in order to understand why certain reactions are occurring.

\subsection{Threats to Validity}
\label{sec:tv}
Due to COVID-19 constraints our testing stage was limited to a single team of 4 participants. The evaluation results obtained are thus preliminary, and further testing is required for improvements drawn from reflections more reflective or real-world use cases to be introduced in our application. Potential bias stemming from the fact that our participants were students undertaking the same research unit also cannot be ruled out.

To address these threats to the validity of our research, additional testing and development will be conducted in the future. These evaluations will be done with software professionals who have had experience working in Agile teams, and ideally with \textit{Emotimonitor} being used in managing a real-world project; it is hoped that testing in a situation more reflective of its intended use case will provide more useful insights regarding its impact and potential improvements.

\subsection{Future Work}
\label{sec:fw}
During the course of these investigations \textit{Emotimonitor} will continue to be refined with an eye towards: improving the user experience for both team members and managers, extending the range of options provided for analyzing and visualising reaction data, and porting \textit{Emotimonitor} to other project management platforms, e.g. JIRA (Atlassian)\footnote{https://www.atlassian.com/software/jira}. 

\section{Related Work}
\label{sec:rw}
Given our motivation to develop a tool to reflect team member emotions, our initial research methodology involved researching existing Trello and JIRA power-ups. This research yielded a dual benefit: it enabled verification of whether our envisioned target tool already existed, and enabled us to examine the advantages and disadvantages of related tools. Ingrained as part of this research was a review of tools currently utilised for emotion-information gathering purposes. Upon closer inspection, it was found that the majority of these tools relied heavily on underlying analytics \citep{werder_meme_2018, thelwall_sentiment_2010, calefato_emotxt_2017} produce contextualized emotional-information. This was in direct opposition to our intended goal of providing users with a direct means of expressing their emotional states, the platform to do so taking the form of emoji reactions. Through our research during this phase of the project, our team was able to overall determine that our envisioned tool did not exist, and thus had potential to greatly benefit Agile teams operating in professional industries without being derivative in nature. Research into how teams respond emotionally to the requirements changes entailed by Agile practices is scarce. Studies on the role of emotions in Agile development have focused on the role of emotional intelligence in Agile teams, from both the management perspective \citep{geoghegan_project_2008} and those of practitioners in general \citep{luong_agile_2019}; although some research indicates that negative emotions can arise from misalignment between project goals and individual goals \citep{cao_understanding_2017}, there is little in the way of concrete approaches or tools recommended to mitigate these risks. The tool we have developed, \textit{Emotimonitor}, aims to better enable managers to monitor the “signs and influence of certain emotions that may lower the team moral [sic]” \citep{cao_understanding_2017}. To the best of our knowledge, tools designed to capture emotional responses during projects have not been widely adopted within industry; existing tools on Trello such as the ‘Card Reaction’ power-up \citep{devenny_card_2020} provide the capacity to display reactions but lack any underlying analytics. In addition, such a tool also lacks an emotion intensity scale in comparison to our intended functionality for users. Nevertheless, existing methods of capturing emotion dynamics primarily focus on extracting emotions from natural language messages and comments \citep{werder_meme_2018, thelwall_sentiment_2010, calefato_emotxt_2017}. A previous approach to capturing and modelling emotion dynamics in projects used one such method, ‘EmoTxt’ \citep{calefato_emotxt_2017}, and focused on the “time dynamic aspects of emotions” \citep{neupane_approach_2019}. \textit{Emotimonitor} differs from this type of approach in that it is more explicit in representing and collecting emotions, using a set of reactions attached to each card through which users express their responses. In addition, when juxtaposing the benefit of alternative tools with \textit{Emotimonitor}, we have provided immediate information to be leveraged by managers and there is no dependency on underlying communication records to identify emotions. 

\section{Conclusion}
In this paper we have presented \textit{Emotimonitor}, a proof-of-concept Trello extension that enables Agile team members to record emotional responses in the face of changing requirements. Unlike existing alternatives that adopt underlying machine learning to produce conclusions of emotion states, \textit{Emotimonitor} provides flexibility for team members to record their emotions real-time through a clear user interface. Accompanying this flexibility is the ability for Agile team managers to gauge the collective emotional states in their team and thereby identify any underlying issues that may be plaguing a team’s productivity. 

\section*{Acknowledgement}
Madampe is supported by a Monash Faculty of IT scholarship, and Grundy is supported by ARC Laureate Fellowship FL190100035.

\section*{CRediT Author Statement}
\textbf{Mohammed-Amr Abd El-Migid, Damon Cai, Thomas Niven, Jeffrey Vo:} Software, Investigation, Writing - Original Draft. \textbf{Kashumi Madampe:} Conceptualization, Methodology, Data Curation, Writing - Review \& Editing, Supervision, Project administration. \textbf{John Grundy:} Conceptualization, Writing - Review \& Editing, Supervision. \textbf{Rashina Hoda:} Conceptualization, Writing - Review \& Editing.

\bibliographystyle{elsarticle-harv} 
\bibliography{main}

\end{document}